\documentclass[pra,twocolumn,floatfix,superscriptaddress]{revtex4}

\usepackage{lscape}
\usepackage{rotating}
\usepackage{epstopdf}
\usepackage{graphicx}
\usepackage{natbib}
\usepackage{epstopdf}
\usepackage{amsmath}
\usepackage{amssymb}
\usepackage{mathbbol}
\usepackage{xcolor}
\usepackage{lipsum}
\usepackage{amssymb}
\usepackage{mwe}
\usepackage{lipsum} 
\usepackage{siunitx}
\usepackage[normalem]{ulem}

\usepackage{empheq}
\usepackage{afterpage}
\usepackage[colorlinks,citecolor=blue,urlcolor=red]{hyperref}
\usepackage{cleveref}

\newcommand{\beq}{\begin{equation}}
\newcommand{\eeq}{\end{equation} \smallskip}
\newcommand{\beqy}{\begin{eqnarray}}
\newcommand{\eeqy}{\end{eqnarray} \smallskip}

\newcommand{\bit}{\begin{itemize}}
\newcommand{\eit}{\end{itemize}}
\newcommand{\bmat}{\begin{pmatrix}}
\newcommand{\emat}{\end{pmatrix}}

\begin{document}

\title{Quench dynamics of an ultracold two-dimensional Bose gas}

\author{P. Comaron}
\thanks{These two authors contributed equally to this work.}
\address{Joint Quantum Centre (JQC) Durham-Newcastle, School of Mathematics, Statistics and Physics,
Newcastle University, Newcastle upon Tyne, NE1 7RU, United Kingdom}

\author{F. Larcher}
\thanks{These two authors contributed equally to this work.}
\address{Joint Quantum Centre (JQC) Durham-Newcastle, School of Mathematics, Statistics and Physics,
Newcastle University, Newcastle upon Tyne, NE1 7RU, United Kingdom}
\address{INO-CNR BEC Center and Dipartimento di Fisica, Universit\`a di Trento, via Sommarive 14, I-38123 Trento, Italy}

\author{F. Dalfovo}
\address{INO-CNR BEC Center and Dipartimento di Fisica, Universit\`a di Trento, via Sommarive 14, I-38123 Trento, Italy} 

\author{N. P. Proukakis}
\address{Joint Quantum Centre (JQC) Durham-Newcastle, School of Mathematics, Statistics and Physics,
Newcastle University, Newcastle upon Tyne, NE1 7RU, United Kingdom}

\date{\today}

\begin{abstract}
We study the dynamics of a two-dimensional Bose gas after an instantaneous quench of an initially ultracold thermal atomic gas across the Berezinskii-Kosterlitz-Thouless phase transition, confirming via stochastic simulations that the system undergoes phase ordering kinetics and fulfills the dynamical scaling hypothesis at late-time dynamics. Specifically, we find in that regime the vortex number decaying in time as $\left< N_\mathrm{v} \right> \propto t^{-1}$, consistent with a dynamical critical exponent $z \approx 2$ for both temperature and interaction quenches.
Focusing on finite-size box-like geometries, we demonstrate that such an observation is within current experimental reach.
\end{abstract}

\maketitle

%%%%%%%%%%%%%%%%%%%%%%%%%%%%%%%%%%%%%%%%%%%

\section{{Introduction}}

A two-dimensional (2D) Bose gas is known to undergo a Berezinskii-Kosterlitz-Thouless (BKT) phase transition \cite{Berezinskii72,Kosterlitz1973} between a superfluid and a non-superfluid regime, originating from the binding and unbinding of vortex-antivortex pairs. Such a transition, first observed in thin helium films \cite{Bishop78}, has been experimentally investigated in a broad range of systems including ultracold atomic gases in quasi-2D harmonic potentials \cite{stock2005,clade2009,hadzibabic2006berezinskii,kruger2007p,Cornell2010,PhysRevLett.107.130401} and, more recently, in near-uniform box-like traps \cite{chomaz2015emergence}. Theoretically such settings have been studied  with a universal $|\Psi|^4$ model on a lattice \cite{Prokofev2001,Prokofev2002}, a semiclassical  field approach \cite{Giorgetti2007},  Quantum Monte Carlo methods \cite{Holzmann2008,Holzmann2010},  classical fields and the stochastic (projected) Gross-Pitaevskii equation applied to both harmonically trapped gases \cite{Simula2006, Simula2008, Bisset2009, Proukakis2012, Mathey2017} and box-like geometries \cite{foster2010vortex,Karl2017,gawryluk2018signatures}, and a renormalization group approach \cite{Pelissetto2013}. In the weakly-interacting regime, such predictions have shown good agreement with experimental findings for the interference fringes in expansion dynamics \cite{hadzibabic2006berezinskii, Polkovnikov2006}, the relation between the number of vortices and the emergence of phase coherence  \cite{chomaz2015emergence}, the scale invariance and universality \cite{hung2011observation, Proukakis2012}, and the propagation of sound near the BKT transition \cite{Ville2018,Ota2018}.

In cold gases, for a given atomic species, the critical temperature of the BKT phase transition is set by the chemical potential and effective interaction strength. These quantities can be experimentally controlled by fixing the atom number and varying either the strength of the transverse confinement or the scattering length  by means of a Feshbach resonance, where available \cite{note-Sakharov}.  The BKT transition can be crossed by lowering the temperature {\it via} evaporating cooling the gas. The dynamics emerging in a temperature quench are sensitive to the quench protocol, and in particular to the quench rate.  A quench across a continuous phase transition spontaneously generates defects in the order parameter. A relation between the defect density and the quench rate is provided by the Kibble-Zurek universal scaling law \cite{kibble1976topology, zurek1985cosmological}. In the case of an instantaneous (or sufficiently rapid) quench, the emerging dynamics reveal a phase-ordering stage in which self-similar correlation functions collapse onto each other when scaled in terms of a characteristic lengthscale \cite{Bray1994,rutenberg1995phase}. The growth of this lengthscale, which is directly connected to the defect dynamics, is set by the dynamical critical exponent $z$ which can thus be extracted from the simulations in the appropriate late-time evolution stage.

The purpose of this work is to theoretically study the phase-ordering kinetics of a quenched ultracold atomic Bose gas in a box trap, within existing and envisaged 2D box geometries, mostly inspired by recent experiments realized in the Laboratoire Kastler Brossel (LKB) in Paris  with $^{87}$Rb \cite{chomaz2015emergence, Ville2018, Saint-Jalm2019} and currently underway at Cambridge (AMOP) with $^{39}$K \cite{Hadzibabic}. We model the gas by means of the stochastic projected Gross-Pitaevskii equation \cite{Stoof2001, Stoof1999, Gardiner2003, Bradley2008,Blakie2008, Proukakis2008}.  Considering the geometry of the LKB box trap, we first perform a detailed analysis of the equilibrium configuration as a function of temperature, with our findings revealing good qualitative agreement with earlier numerical works.  Having identified the relevant regimes, we then discuss controlled instantaneous quenches across the BKT phase transition. Firstly, we verify the expected bulk predictions by considering both temperature and interaction quenches in the limit of a large `idealized' box -- larger than currently accessible experimentally. Having demonstrated our methodology and confirmed its predictive power both in terms of correlation functions and vortex dynamics, we then specifically address the feasibility of experimental observation.
Correlation functions -- which are hard to measure directly in 2D box geometries -- are likely to be prone to finite size effects in the currently-accessible box sizes, suggesting it may be favorable for observations to rely instead on vortex numbers.
Considering the specific finite extent of the recent LKB experimental set-up in Paris \cite{Ville2018}, we show that even in such finite-size systems, there is a well-defined temporal window in which one should be able to observe the long-term $t^{-1}$ evolution in the vortex number, associated with vortex-antivortex annihilation processes.  We show that the same conclusion is valid also in the different setting inspired by ongoing experiments of AMOP in Cambridge. Similar findings have been previously obtained in the long-term evolution of the closely-related problem of decaying two-dimensional quantum turbulence, connecting ultracold atom experiments~\cite{Kwon2014,Kim2016,Seo2017,Neely2013} with numerical studies \cite{Baggaley2018}, a problem discussed more generally in the context of dynamical vortex decay via $N$-vortex collision processes \cite{Baggaley2018,Groszek2016,Groszek2019,Karl2017}.

%%%%%%%%%%%%%%%%%%%%%%%%%%%%%%%%%%%%%%%%%%%

\section{Stochastic Gross-Pitaevskii Equation}
\label{sec:coherent_pump}

We consider a weakly interacting ultracold Bose gas confined in a transverse tight potential sufficiently strong that it creates a uniform 2D system which occupies the lowest energy state in the transverse direction. The stochastic (projected) Gross-Pitaevskii equation (SPGPE) describes its dynamics via the noisy complex field $\psi(x,y,t)$, subject to the equation  \cite{Stoof2001, Stoof1999, Gardiner2003,Bradley2008, Blakie2008, Proukakis2008}
\begin{multline}
\label{eq:SPGPE}
i \hbar \frac{\partial \psi(x,y,t)}{\partial t} =\hat{\mathcal{P}}\left\{(1-i\gamma )\left[-\frac{\hbar^2 \nabla^2}{2m_p} + \right.  \right.   \\
\left.  g_\textrm{2D}|\psi(x,y,t)|^2-\mu \bigg]{\psi(x,y,t)} +\eta(x,y,t)\right\} \, ,
\end{multline}
where $\nabla^2$ is the Laplacian in two dimensions and $g_\textrm{2D}$ is the 2D coupling constant associated with the \emph{s}-wave scattering length $a_s$, and $m_p$ is the mass of the particles. If the transverse confinement is harmonic, $V(z)=m_p \omega_z^2/2$,  one can define the harmonic length  $\ell_\perp=\sqrt{\hbar/m_p\omega_z}$ and the dimensionless coupling constant $\tilde{g} = m_p g_\textrm{2D} / \hbar^2 = \sqrt{8\pi}{a_s}/{\ell_\perp}$. The projector $\hat{\mathcal{P}}$ constrains the dynamics of the system within a finite number of macroscopically occupied modes, the \emph{coherent region}, up to an ultraviolet energy cutoff fixed here as
\begin{equation}
	\label{eq:cutoff}
	\epsilon_\mathrm{cut} (\mu,T) =k_\mathrm{B}T\log(2)+\mu \, ,
\end{equation}
for which the mean occupation of the last included mode is of order $\sim 1$, assuming a Bose-Einstein distribution in the occupation number spectrum. {A Gaussian stochastic noise $\eta(x,y,t)$ with correlation in space and time $\langle \eta^*(x,y,t)\eta(x',y',t')\rangle = 2\hbar\gamma k_\mathrm{B}T\delta(x-x')\delta(y-y')\delta(t-t')$ is added and projected in the coherent region. The cutoff is implemented in Fourier space. } 
The numerical grid for the simulations is determined from $\epsilon_\mathrm{cut}$ by the anti-aliasing condition $\Delta x \leq \pi/\sqrt{8m\epsilon_\mathrm{cut}}$ \cite{Blakie2008}. In comparing numerical results with experiments, one must include the density of atoms in the coherent region $n_{c\textrm{-field}}=\int dxdy\,|\psi(x,y)|^2$ as well as in the incoherent region $n_\mathrm{I}$, so that the total density is $n=n_{c\textrm{-field}}+n_\mathrm{I}$. 
{Here $n_\mathrm{I}$ is determined by the density of states $G(\epsilon)$ of the system as $n_\mathrm{I} =1/(L_xL_y)\int_{\epsilon_\mathrm{cut}}^\infty d\epsilon\, G(\epsilon)/(e^{(\epsilon-\mu)/k_\mathrm{B}T}-1)$, where the gas in the incoherent region is assumed to be ideal.}
Finally, the dissipative term $\gamma$, parametrizing the interaction between the high-lying and the low-lying modes of the system, has the practical role of setting the rate at which the system reaches the equilibrium determined by the external parameters temperature $T$ and chemical potential $\mu$. In our  simulations we use values of $\gamma$ within a decade centered in the value of $\gamma=0.01$, of the same order of the values used in  \cite{liu2018dynamical} where $\gamma$ was fixed to reproduce the time growth of the number of atoms occupying the lowest-momentum mode in a 3D condensate subject to a temperature quench across the BEC transition \cite{Donadello2016}.

\begin{figure*}[t]
	\centering
		\includegraphics[width=1.32\columnwidth]{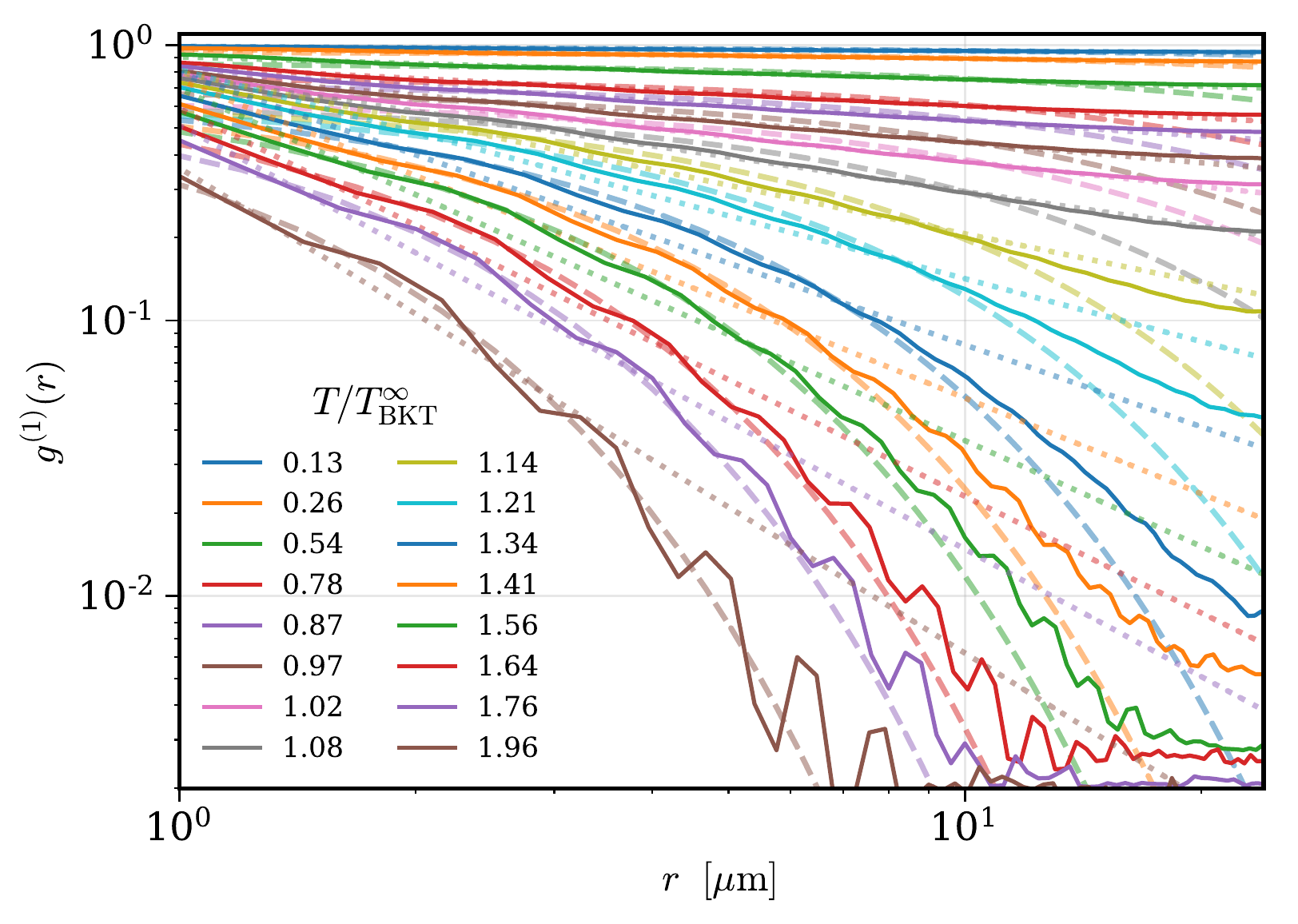}
		\includegraphics[width=0.69\columnwidth]{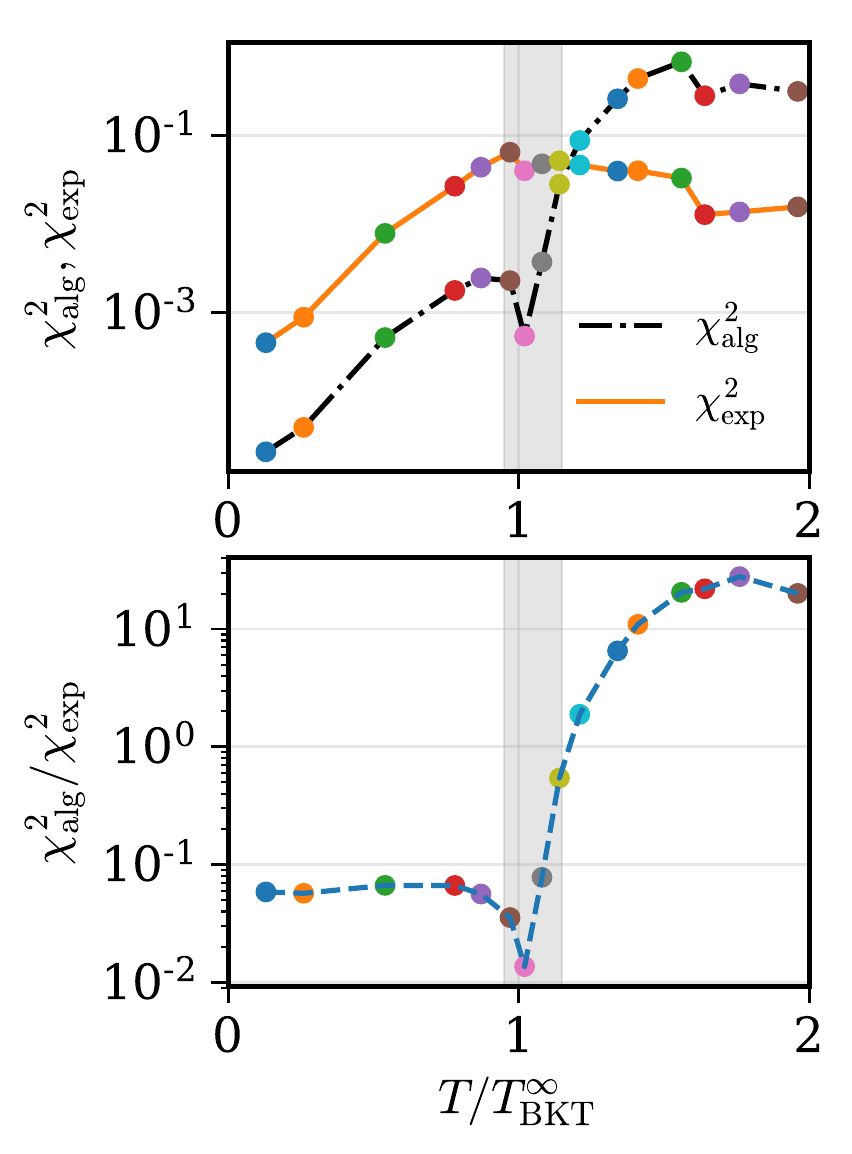}
	\caption{Left panel: equilibrium profiles for the first-order correlation function $g^{(1)}(r)$ for different temperatures, in logarithmic scale. The fitted algebraic (dotted lines) and exponential (dashed lines) functions are defined as in Eq. \eqref{eq:gonefitrel}. The numerical box has size $L_x \times L_y = (50 \times 50)\mu$m, with periodic boundary conditions, and contains $20,000$ $^{87}$Rb atoms. The averages are performed over $\mathcal{N} = 100$ stochastic realizations for each temperature data set and we used $\gamma=0.05$.
	Right panel: $\chi^2$ of the exponential and the algebraic fits (top) and their ratio (bottom) as a function of $T/T_\mathrm{BKT}^{\infty}$. 
	The colored points correspond to the colored lines in the left panel; the critical region is reported as a shaded area.
	}
	\label{fig:gonefit}
\end{figure*}

%%%%%%%%%%%%%%%%%%%%%%%%%%%%%%%%%%%%%%%%%%%

\section{{Equilibrium: the Berezinskii-Kosterlitz-Thouless phase transition}} 

\subsection{Background Theory}

The Mermin-Wagner-Hohenberg theorem \cite{Mermin1966,Hohenberg1967} states that, for a system of dimensions 2 or fewer and short range interactions, it is not possible to have the spontaneous breaking of a continuous symmetry at any non-zero temperature. As a notable consequence, there is no Bose-Einstein condensation in these geometries, since the thermal fluctuations at any temperature are strong enough to destroy the long-range coherence in the system. Repulsive interactions, however, may result in the establishment of quasi-long range coherence at sufficiently low temperatures, affecting the behavior of the first-order correlation function defined in an isotropic system as
\begin{equation}
\label{eq:gonedef}
g^{(1)}(r) = \frac{\langle \psi^*(\vec{r}_0)\psi(\vec{r}_0+\vec{r})\rangle_{\vec{r}_0,\theta,\mathcal{N} }}{\sqrt{\langle |\psi(\vec{r}_0)|^2\rangle_{\vec{r}_0,\theta,\mathcal{N} } \langle |\psi(\vec{r}_0+\vec{r})|^2\rangle_{\vec{r}_0,\theta,\mathcal{N} }}} \, ,
\end{equation}
where the average $\langle \dots \rangle_{\vec{r}_0,\theta,\mathcal{N} }$ is performed over the spatial position $\vec{r}_0$, the angular part $\theta$ of $\vec{r}$ and over a large number $\mathcal{N}$ of stochastic realizations. In a 2D system, as shown in  \cite{Kosterlitz1973}, the topological properties of the system are determined by the behavior of quantized vortex pairs with respect to a critical temperature $T_\mathrm{BKT}$ for the infinite order Berezinskii-Kosterlitz-Thouless phase transition. Specifically:
\begin{itemize}
	\item for $T>T_\mathrm{BKT}$ free vortices can exist in the system, there is no superfluid and the first-order correlation function decays as $g^{(1)}(r)\sim e^{-r/\xi}$, where $\xi$ is a correlation length;
	\item for $T<T_\mathrm{BKT}$ vortices can only exist in bound pairs, allowing the presence of a superfluid, and the correlation shows an algebraic decay $g^{(1)}(r)\sim r^{-\alpha}$ in terms of an exponent $\alpha$.
\end{itemize}
Note that the algebraic decay of the correlation function in the degenerate case could lead to very strong requirements in terms of the sample size, for behavior consistent with the Mermin-Wagner-Hohenberg theorem to manifest itself. At the transition ($T=T_\mathrm{BKT}$), the correlation function should decay according to \cite{Kosterlitz1973,kosterlitz1974critical}
\begin{equation}
\label{eq:gonecrit}
g^{(1)}(r)|_c=\left(\frac{r}{\lambda_T}\right)^{- \alpha_\mathrm{c}},
\end{equation}
where $\lambda_T=\sqrt{2\pi \hbar^2/{m_p} k_\mathrm{B}T}$ is the thermal de Broglie wavelength,
and $\alpha_\mathrm{c}= 0.25$ \cite{Nelson1977}.
At the thermodynamic limit, when the volume and total number of particles of the system tend to infinity while the density is fixed,
the value for the critical temperature is determined by
\begin{equation}
\label{eq:g_crit}
\frac{\mu}{k_\mathrm{B}T} \bigg\rvert_\mathrm{BKT} \approx \frac{\tilde{g}}{\pi} \ln\left(\frac{C}{\tilde{g}}\right),
\end{equation}
where the constant $C$ has been estimated by Monte-Carlo analysis in \cite{Prokofev2002} to be $C\sim 13.2$.
Thus, by inverting Eq.~\eqref{eq:g_crit}, the temperature at the critical point is estimated as
\begin{equation}
\label{eq:Tbkt_inf}
T_{\mathrm{BKT}}^{\infty} =  \frac{{\hbar^2} \pi n}{ {m_p k_\mathrm{B}} \ln{\left( C / \tilde{g} \right)}} ,
\end{equation}
where $n$ denotes the system density.

%%%%%%%%%%%%%%%%%%%%%%%%%%%%%%%%%%%%%%%%%%%

\subsection{Numerical Results}

We now characterize the equilibrium state of a finite-size uniform 2D Bose gas, focusing on the experimental geometry at LKB. We consider approximately 20,000 $^{87}$Rb atoms with mass ${m_p} = 1.4431 \times 10^{-25} \si{\kilogram}$ in a uniform two-dimensional box of size $L_x\times L_y = (50 \times 50) \mu$m, whose transverse confinement is $\omega_z = 2\pi ( 1500 ) \textrm{Hz}$. The chemical potential is fixed at $\mu/k_\textrm{B} = 4.8 \textrm{nK}$, and  $\tilde{g} = 9.5 \times 10^{-2}$. For our choice of parameters, the system is numerically equilibrated for a time $t_\mathrm{eq}=3\times 10^5\omega_z^{-1} \approx 30$s  at a temperature spanning an interval $T/T_\mathrm{BKT}^{\infty} = [0.13,1.96]$, where $T_{\mathrm{BKT}}^{\infty} = 33.25 \mathrm{nK}$. Our simulations were performed by means of the XMDS2 software package described in~\cite{XMDS2013}, on the High-Performance-Computing cluster at Newcastle University. 

To obtain a smooth first-order correlation function, $g^{(1)}(r)$, we average over both $x$ and $y$ directions in our discretized grid (of $N_x \times N_y$ points, respectively) using the expression
\begin{equation}
g^{(1)}(r) \equiv \frac{1}{2}  [ g_x^{(1)}(r) + g_y^{(1)}(r) ]
\end{equation}
where 
\begin{equation}
\label{eq:gonediscr}
\begin{alignedat}{2}
&g_x^{(1)}(r) &= \frac{1}{N_{x}N_y} \sum_{i=1}^{N_{x}}\sum_{j=1}^{N_{y}} \frac{\langle\psi^*_{i,j}\psi_{i+r,j}\rangle_{\mathcal{N}}}{\sqrt{\langle |\psi_{i,j}|^2\rangle_{\mathcal{N}} \langle |\psi_{i+r,j}|^2\rangle_{\mathcal{N}}}} \\
&g_y^{(1)}(r) &= \frac{1}{N_{x}N_y} \sum_{i=1}^{N_{x}}\sum_{j=1}^{N_{y}} \frac{\langle\psi^*_{i,j}\psi_{i,j+r}\rangle_{\mathcal{N}}}{\sqrt{\langle |\psi_{i,j}|^2\rangle_{\mathcal{N}} \langle |\psi_{i,j+r}|^2\rangle_{\mathcal{N}}}} \, ,
\end{alignedat}
\end{equation}
where $\psi_{i,j} = \psi(x_i,y_j)$ and the average $\langle \dots \rangle_\mathcal{N}$ is performed over the number $\mathcal{N}$ of stochastic realizations.
Once this function is computed, we fit it  (as in \cite{Dagvadorj2015,Comaron2018}) with the functions
\begin{equation}
g^{(1)}_\text{exp}(r) \propto e^{-r/\xi}  \quad \text{and} \quad g^{(1)}_\text{alg}(r) \propto r^{-\alpha} \, .
\label{eq:gonefitrel}
\end{equation}
According to the BKT theory, the former should apply above $T_{\mathrm{BKT}}$ and the latter below. As a measure of the quality of the fit we use the quantity 
\begin{equation}
\chi^2_\mathrm{fit}=\sum_i \frac{(g^{(1)} (r_i) - g^{(1)}_\mathrm{fit}(r_i))^2}{(g^{(1)}(r_i))^2},
\end{equation}
where the index $i$ accounts again for the spatial discretization. 

The results are given in Fig.~\ref{fig:gonefit}. The equilibrium correlation functions and the corresponding fits are shown in the left panel for the selected temperature range, while the $\chi^2$ functions are shown on the right. As expected, the algebraic fit is better than the exponential fit (i.e., has a lower $\chi^2$) at low temperatures, while the exponential fit is better at high temperatures. The crossover between the two behaviors occurs in a narrow region close to $T_{\mathrm{BKT}}^{\infty}$, the two values $\chi^2_{\rm exp}$ and $\chi^2_{\rm alg}$ being equal slightly above $T_{\mathrm{BKT}}^{\infty}$. The shaded area between $ 0.95 < T/T_\mathrm{BKT}^{\infty} < 1.15$ represents qualitatively the region where the BKT transition occurs. 

\begin{figure}[t]
	\centering
	\includegraphics[width=\columnwidth]{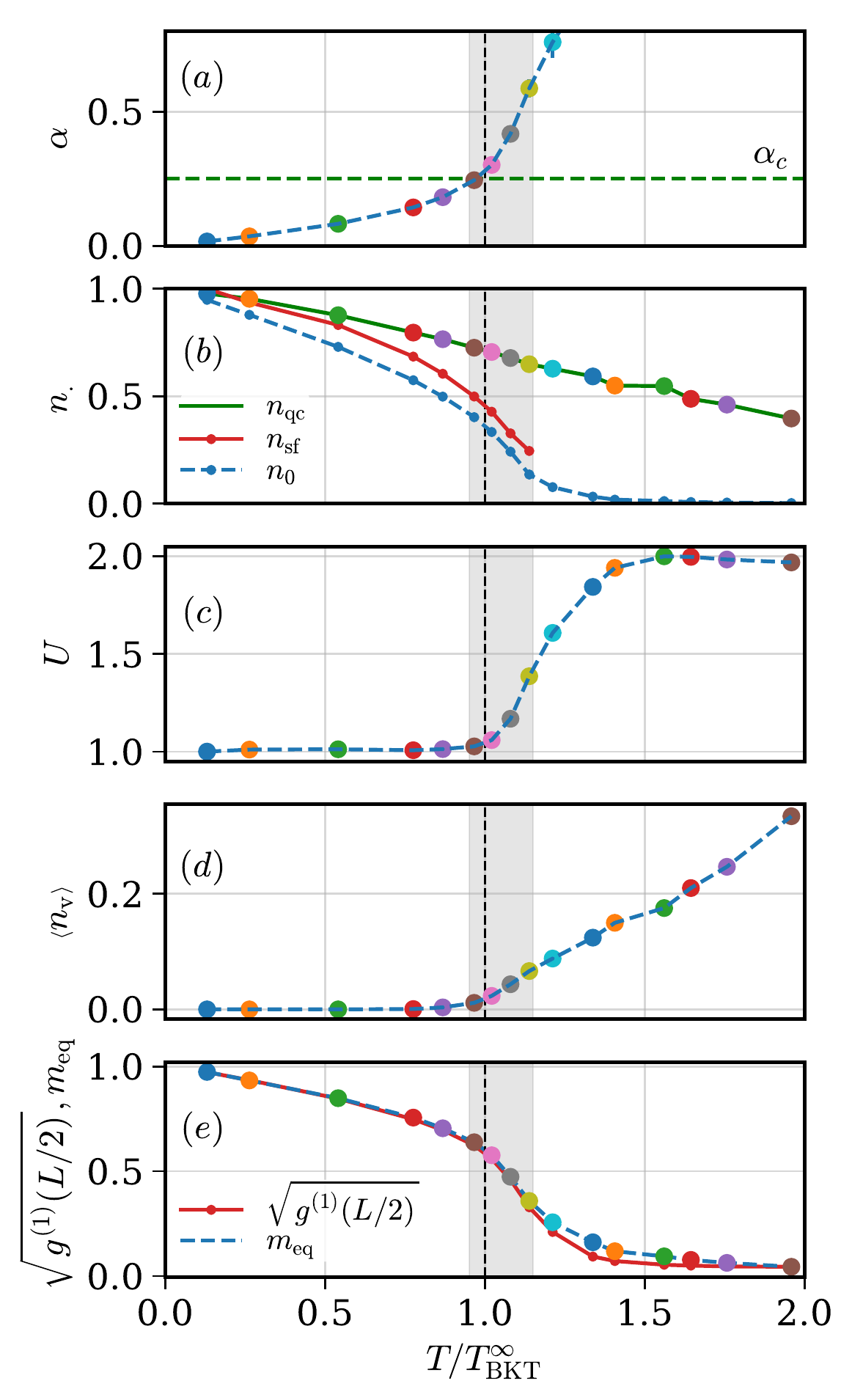}
	\caption{
		(a) Exponent $\alpha$ of the algebraic fit as in Eq.~\eqref{eq:gonefitrel}, calculated
		for samples at thermal equilibrium at different values of $T/T_\mathrm{BKT}^{\infty}$ as in Fig.~\ref{fig:gonefit}.
		{The critical value $\alpha_c=0.25$ predicted by BKT theory in the thermodynamic limit is shown as a horizontal dashed green line.}
		(b) Equilibrium quasi-condensate fraction $n_{\mathrm{qc}}$  (green solid line and coloured {big} dots), superfluid fraction $n_\mathrm{sf}$ (red {small} dots and solid line) and condensate fraction $n_{0}$ (blue {small} dots and {dashed} line) as a function of temperature.
		(c) Equilibrium Binder ratio $U$ as defined in {Eq.~\eqref{eq:binder}} for the same samples.
		(d) Equilibrium vortex density computed by performing a short time
		average of the final values of $\left< N_\mathrm{v} \right>$. 
		(e) Equilibrium order parameter $m$ as defined in Eq.~\eqref{eq:orderpar}, in comparison with the equilibrium value of $\sqrt{g_1(r)}$ computed at the edges of the system. 
	}
	\label{fig:multiplot}
\end{figure}

In the top panel of Fig.~\ref{fig:multiplot} we show the parameter $\alpha$ extracted from the algebraic fit. The shaded area is the same as in Fig.~\ref{fig:gonefit}. At the left border of this area we find $\alpha \approx 0.25$, which corresponds to the prediction of BKT theory at the transition in the thermodynamic limit~\cite{kosterlitz1974critical,Nelson1977} and in agreement with \cite{Nazarenko2014}.  At the right border we instead find $\alpha \approx 0.5$. This value agrees with the one obtained in \cite{foster2010vortex} just above $T_{\mathrm{BKT}}$ in numerical simulations within a classical field approach in the micro-canonical ensemble, reproducing the interference patterns experimentally observed in \cite{hadzibabic2006berezinskii}. A broad critical region, as identified here, is also consistent with the existence of  an ``intermediate regime" characterized by $\alpha > 0.25$ and a rapidly vanishing zero momentum current-current correlation, as pointed out in \cite{gawryluk2018signatures}, again within a classical field calculation but in the grand-canonical ensemble.  Such a region is predicted to shrink when approaching the thermodynamic limit \cite{foster2010vortex,gawryluk2018signatures}.

Although no true condensation can occur in 2D in the presence of interactions, the quasi-long range coherence leads to the formation of a so called ``quasi-condensate", which can be thought of as a condensate with a fluctuating phase \cite{Kagan1987}. Its density can be computed via \cite{Prokofev2002}
\begin{equation}
\label{eq:quasicond}
n_\mathrm{qc} = \frac{\sqrt{2\langle|\psi(\vec{r})|^2\rangle_{\vec{r},\mathcal{N} }^2-\langle |\psi(\vec{r})|^4\rangle_{\vec{r},\mathcal{N}}}}{n},
\end{equation}
where the average of the moduli $\langle \dots \rangle_{\vec{r},\mathcal{N} }$ is performed over the entire spatial grid and all noise realizations. As shown by the green line in Fig.~\ref{fig:multiplot}(b) -- and consistent with earlier works \cite{foster2010vortex} -- the quasi-condensate density has a significant non-zero value even above the critical temperature, with $n_\mathrm{qc}(T_\mathrm{BKT}^{\infty})\sim 0.7$. In the framework of the BKT theory, the quasi-condensate facilitates the existence of vortices above the critical temperature, where a superfluid is absent. In the same panel of Fig.~\ref{fig:multiplot} we also report the normalized superfluid density $n_\mathrm{sf}$ (red line), defined as
\begin{equation}
n_\textrm{sf}  = \frac{1}{\lambda_T^2 \alpha},
\end{equation}
as suggested by Nelson and Kosterlitz \cite{Nelson1977}, which is meaningful only at low $T$ where the algebraic fit to $g^{(1)}$ is reliable. We also plot the condensate fraction $n_{0}$ (blue dashed line), defined as the normalized density of particles which populate the zero momentum ($k=0$) mode. At low $T$, the lowest mode tends to saturate at the same value as $n_{\rm qc}$, while above $T_\mathrm{BKT}$ it is just a small fraction, as expected for a strongly fluctuating quasi-condensate.  We note in passing that in our simulations the $c$-field fraction lies between $0.84 < n_{c\textrm{-field}}/n < 0.99$ depending on the temperature. Here, the $c$-field density is calculated as $n_{c\textrm{-field}} = \langle \sum_{i,j} \left| \psi_{i,j} \right|^2\rangle_{\mathcal{N}}$. At the critical point we have $n_{c\textrm{-field}}/n(T_\mathrm{BKT}^{\infty}) = 0.92$.

A related quantity characterizing the location of the phase transition is the Binder ratio (or Binder cumulant) \cite{binder1981,cugliandolo2016}, defined as
\begin{equation}
U = \frac{\langle \big|\sum_{i,j}\psi_{i,j}\big|^4 \rangle_{\mathcal{N}}}{\langle \big|\sum_{i,j}\psi_{i,j}|^2 \rangle_{\mathcal{N}}^2},
\label{eq:binder}
\end{equation}
and plotted in Fig.~\ref{fig:multiplot}(c). This quantity is predicted to be a step-function from $1$ (fully coherent system) to $2$ (pure thermal state) in the limit of infinitely large boxes. In finite volumes, it is instead particularly sensitive to finite-size effects, resulting in a smooth function with a slope progressively small with smaller boxes. 

Another way to characterize the physical regime of the system is by counting the average number of vortices at equilibrium as a function of temperature; this is done by means of a numerical routine that calculates the phase winding around each grid point, to identify the vortices and their circulation. The results are given in Fig.~\ref{fig:multiplot}(d). Note that, for the parameters of our simulations, all vortices annihilate for temperatures lower than  $\sim 0.75 T_\mathrm{BKT}^{\infty}$.

Finally, a measure of the degree of degeneracy of the system was introduced in Ref.~\cite{cugliandolo2016} in the form of an order parameter $m$ which, in our discretized space, is defined as 
\begin{equation}
\label{eq:orderpar}
m = \frac{1}{\sqrt{N_x N_y}} \frac{\langle \left| \sum_{i,j} \psi_{i,j} \right| \rangle_{\mathcal{N}}}{\langle \sum_{i,j} \left| \psi_{i,j} \right|^2\rangle_{\mathcal{N}}} \, .
\end{equation}
This quantity accounts for the off-diagonal terms of the system's density matrix. Its value  at equilibrium, $m_\mathrm{eq}$,  is plotted in Fig.~\ref{fig:multiplot}(e)  together with the value of $\sqrt{g_1(r)}$ calculated at the edges of the system. These quantities should coincide in the limit of large boxes~\cite{cugliandolo2016}. The behaviour of $m_\mathrm{eq}$ is qualitatively similar to the one observed in \cite{cugliandolo2016} for a planar 2D $XY$ model, showing a rapid decrease above the critical temperature. 

All quantities plotted in Fig.~\ref{fig:multiplot} demonstrate the occurrence of the BKT transition within a critical range of $T$. Due to finite-size effects the transition is not sharp. Depending on the way one conventionally defines a precise value of the transition temperature $T_\mathrm{BKT}$, such a value can be slightly shifted downward, as in \cite{gawryluk2018signatures}, or upward, as in \cite{foster2010vortex}, with respect to the ideal value $T_\mathrm{BKT}^{\infty}$, but the different definitions are expected to converge in the thermodynamic limit. Our results demonstrate clear consistency with such earlier findings, as well as with the results of similar analysis in exciton-polariton 2D gases \cite{Comaron2018,furthertest}.

%%%%%%%%%%%%%%%%%%%%%%%%%%%%%%%%%%%%%%%%%%%%%%%%%%%%%%%%%%%%%%%%%%
%%%%%%%%%%%%%%%%%%%%%%%%%%%%%%%%%%%%%%%%%%%%%%%%%%%%%%%%%%%%%%%%%%

\begin{figure*}[]
	\centering
	\includegraphics[width=\textwidth]{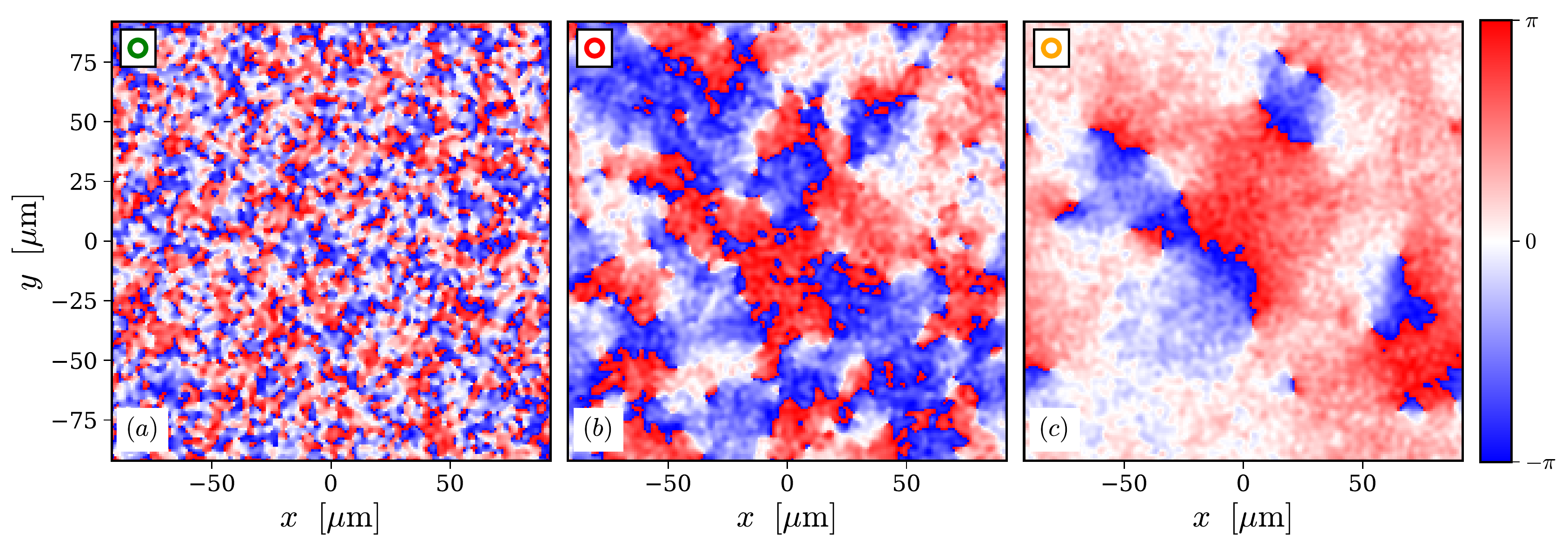}
	\hspace{0.2cm}
	\includegraphics[width=0.94\columnwidth]{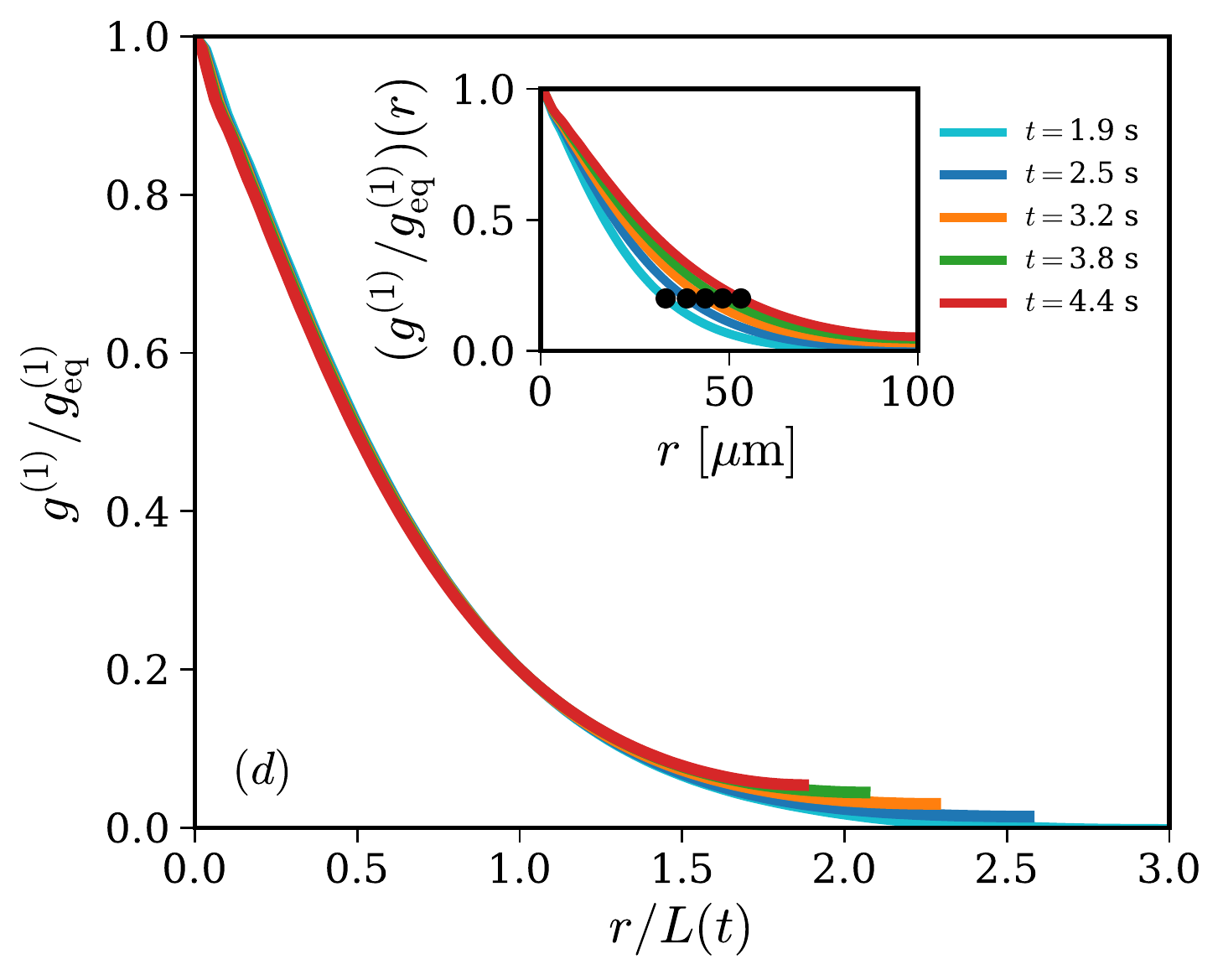}
	\includegraphics[width=1.1\columnwidth]{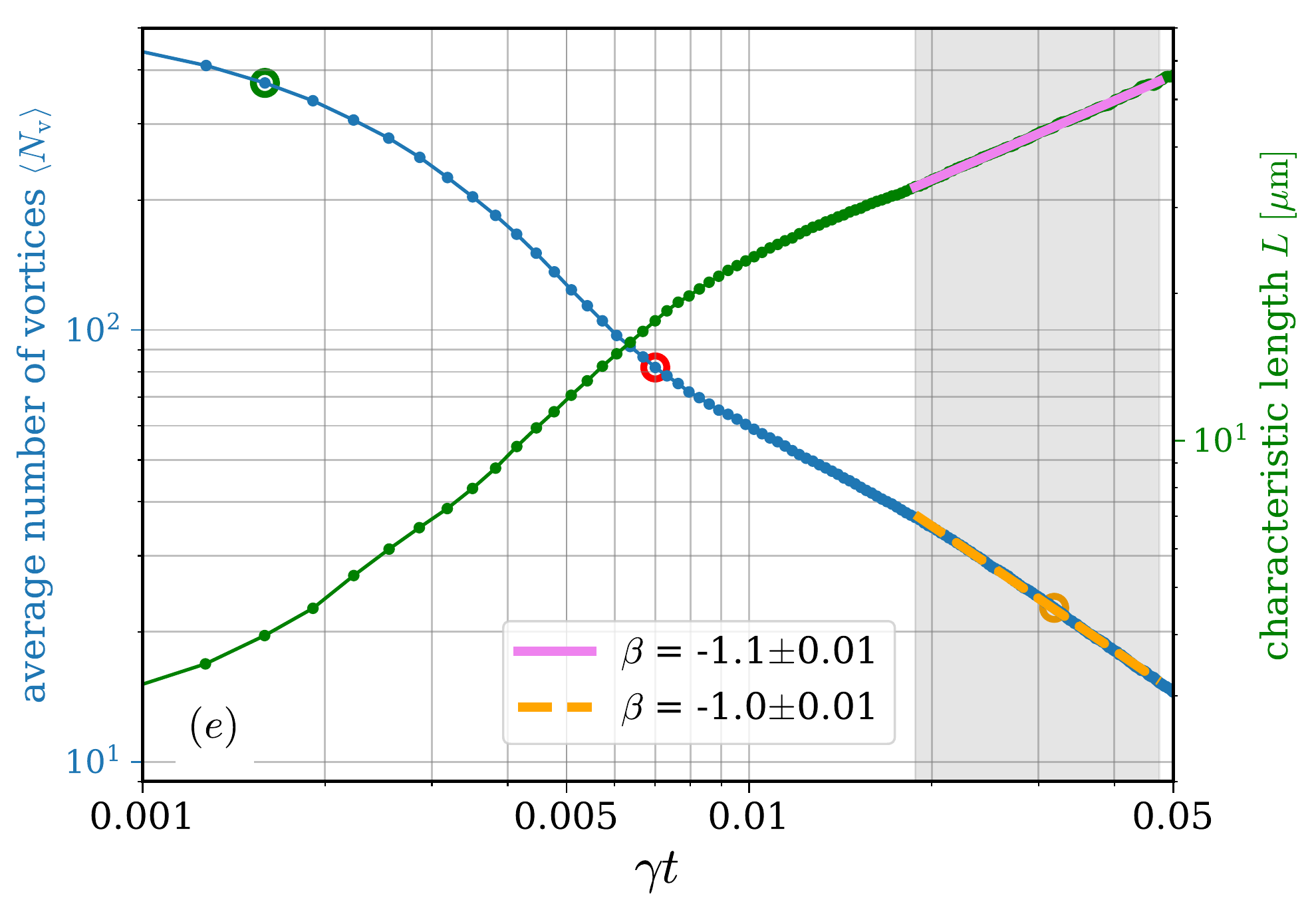}	
	\hspace{0.2cm}
	\includegraphics[width=0.94\columnwidth]{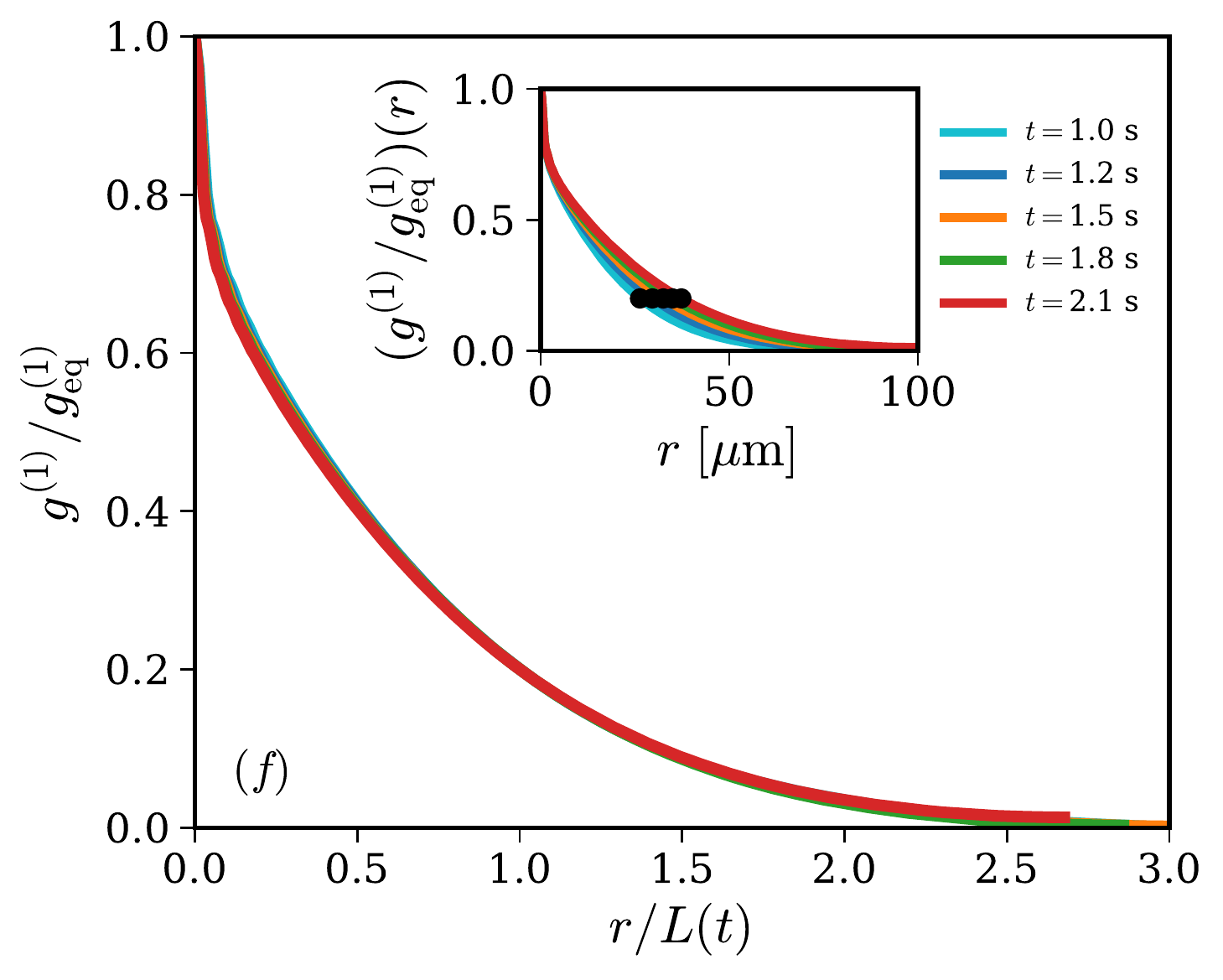}
	\includegraphics[width=1.1\columnwidth]{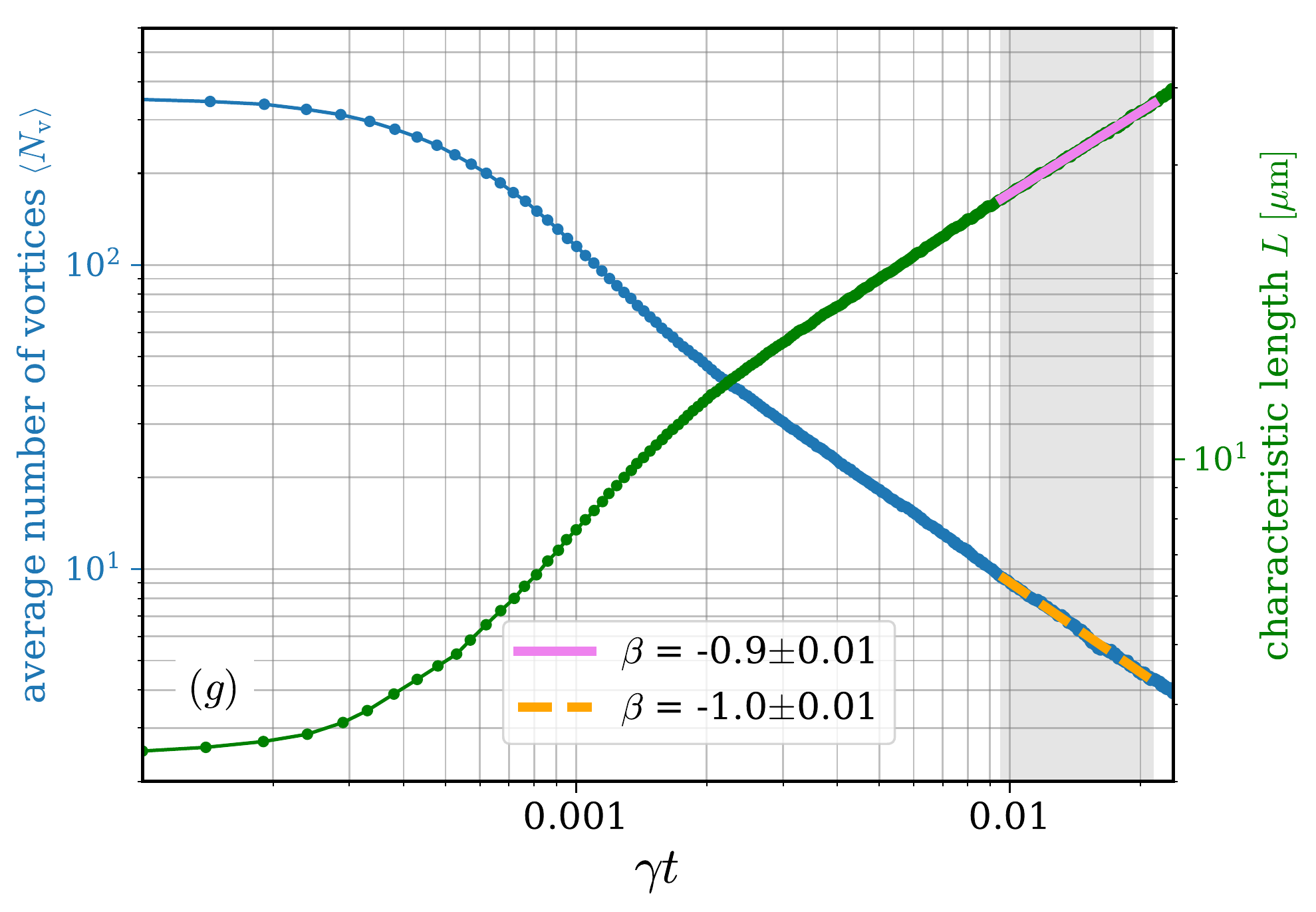}	
	\caption{
	Phase distribution before (a), close to (b) and far after (c) the critical point in a single realization of an instantaneous temperature quench across the BKT transition in a $(200\mathrm{\mu m} \times 200\mathrm{\mu m})$ box with periodic boundary conditions.
	(d) Normalized correlation function $g^{(1)}/g^{(1)}_\mathrm{eq}$ as a function of $r$ at different times (inset).  The value of $r$ where this quantity is equal to $0.2$ (black dots) is used to define the correlation length $L(t)$, which is then used to plot $g^{(1)}/g^{(1)}_\mathrm{eq}$ as a function of the rescaled distance $r/L(t)$. 
	(e) Temporal evolution of number of vortices (blue{, decreasing}) and correlation length $L(t)$ (green, {increasing}) averaged over $\mathcal{N} = 400$ stochastic realizations of the same temperature quench. 
	(f),(g) Same as in the central row but for an instantaneous quench of the interaction parameter $\tilde{g}$, at fixed temperature.  
	For both quench protocols we obtain a scaling exponent $\beta \approx -1$ within the gray-shaded regions.
	In  all simulations we have used $\gamma = 0.01$; for this choice of $\gamma$, the shaded region in the center-right (bottom-right) figure corresponds to $ 1.9 < t < 4.8 \textrm{s}$ ($ 1.0 < t <2.1 \textrm{s}$).
		}
	\label{fig:ideal-box-protocol}
\end{figure*}

%%%%%%%%%%%%%%%%%%%%%%%%%%%%%%%%%%%%%%%%%%%

\section{{Quench and phase ordering dynamics}} 
\label{sec:quench_dynamics}

An important part of our analysis is devoted to the study of the relaxation dynamics of coherence and topological defects 
of the 2D Bose gas across the critical temperature.
This study can reveal universal properties through the characterization of critical exponents, which
address the universality class of the system considered \cite{Bray1994}.
Such a process has been largely studied in conservative \cite{Damle1996,Karl2017} and open systems \cite{Kulczykowski2017,Comaron2018}.
The 2D Bose gas is known to belong to the same universality class of the planar 2D $XY$ model, whose dynamics has been studied theoretically \cite{jelic2011quench} and experimentally \cite{Nagaya1992,Nagaya1995}. 

In Eq.~\eqref{eq:SPGPE}, the parameters controlling the physical state of a given atomic species are $T$, $\mu$ and $g_\mathrm{2D}$. In theoretical studies assuming a fixed value for the interaction strength $g_\mathrm{2D}$, quenching through the critical point of the transition has been implemented by manipulating the temperature parameter $T$ while keeping the chemical potential unaltered \cite{weiler2008spontaneous, jelic2011quench}, by quenching $\mu$ maintaining a fixed temperature \cite{damski2010soliton}, and by performing a simultaneous quench of the two quantities \cite{Liu2016,liu2018dynamical}. 

%%%%%%%%%%%%%%%%%%%%%%%%%%%%%%%%%%%%%%%%%%%

\subsection{Background Theory}

We start by briefly reviewing the theory of the phase ordering process.
The quantities which characterize the nonequilibrium dynamics of a 2D Bose gas are the correlation length $L(t)$, the average length within which coherence is established, and the average number of topological defects detected when crossing the phase transition. In our case, the defects are quantized vortices and their average number at a time $t$ relates to $L$ as
\begin{equation}
\left< N_\textrm{v} \right>^{-1/2}  {\propto}\  L \, ,
\label{eq:relation_L_nv}
\end{equation}
and the corresponding density is $\left<n_\mathrm{v} \right>= \left< N_\mathrm{v} \right> / L_\mathrm{box}^2$,
where from now on $\left< \dots \right>$ corresponds to an average over the stochastic realizations.
After a sudden quench from a temperature above $T_\mathrm{BKT}$, 
at sufficiently long times, the system is expected to enter a temporal region
where the system exhibits \textit{universal dynamical scaling} \cite{Yurke1993,Bray1994}, 
in the sense that the (non-equilibrium) correlation function \eqref{eq:gonedef} evolves in time according to  the dynamical scaling form \cite{Bray1994}
\begin{equation}
	g^{(1)} ({r},t) \sim g^{(1)}_\mathrm{eq}({r}) F\left( \frac{r}{L(t)}  \right)\; ,
	\label{eq:corre_scaling}
\end{equation}
where $g^{(1)}_\mathrm{eq}({r})$ is the static (equilibrium) correlation function for the final parameters $\mu, T$ and $\tilde{g}$, decaying algebraically at long distances as $g^{(1)}_\mathrm{eq}(r) \propto r^{-\alpha}$. 
From this scaling law one can extract the correlation length $L(t)$, which 
is predicted to exhibit a power law dependence on the inverse of time, whose exponent is referred as the \textit{dynamical exponent} $z$.
Accordingly, the  average number of vortices is expected to behave as
\begin{equation}
\left< N_\textrm{v} (t) \right>  {\propto}  \  t^{\beta},
\end{equation}
where $\beta = -2/z$. The dynamical exponent for the diffusive dynamics of the 2D $XY$ model is predicted to be $z \approx 2$ ~\cite{jelic2011quench}.
Note that due to the presence of steady state vortices, logarithmic corrections to this law are also expected, such that $\left< N_\textrm{v} (t) \right>  \propto \ \left( {t}/{\ln(t/t_0)}  \right)^{\beta}$, where $t_0$ is a nonuniversal temperature-dependent timescale (we refer to Refs.~\cite{jelic2011quench,Comaron2018} for details). 
As logarithmic corrections only manifest themselves in very large systems and are unlikely to be directly detected in ultracold atomic experiments, we do not focus on such corrections here, simply noting that the numerical results presented here are in principle consistent with the presence of weak logarithmic corrections.

%%%%%%%%%%%%%%%%%%%%%%%%%%%%%%%%%%%%%%%%%%%

\subsection{Results for a large box}
\label{sec:ideal_box}

In this section we discuss the rather idealized case of a large 2D box trap, to reveal the key physics expected in the homogeneous thermodynamic limit. Specifically we simulate here a $(200 \times 200) \mathrm{\mu m}$ box with periodic boundary conditions, following an instantaneous quench from above to below the BKT phase transition, as a grand canonical evolution.

Firstly, we simulate an infinitely-rapid temperature quench across the critical point of a $^{87}\mathrm{Rb}$ gas by evolving Eq.~\eqref{eq:SPGPE} from an equilibrium initial configuration above the critical point to a quasi-ordered state.
Similarly to earlier works \cite{Liu2016,liu2018dynamical}, we induce a simultaneous jump in the system chemical potential from $\mu <0$ to symmetrically-located (about zero) $\mu > 0 $ values.
Specifically,  we prepare our system in an equilibrated disordered state with temperature $T_\mathrm{in} = 200 \ \mathrm{nK}\gg T_{\mathrm{BKT}}$ and $\mu_\mathrm{in} = -2.4 \ k_\mathrm{B} \ \mathrm{nK} < 0$ and induce at $t=0$ a sudden quench by setting values for a chosen final state with $T_\mathrm{fin} = 5 \ \mathrm{nK} \ll T_{\mathrm{BKT}}$ and $\mu_\mathrm{fin} = 2.4 \ k_\mathrm{B} \ \mathrm{nK} >0$. 
The chosen values of $\mu$ and $T$ also set the cutoff for each stage of the system evolution.
Interaction strength $\tilde{g} = 95 \times 10^{-3}$ is fixed by the transverse confinement adopted, namely $\omega_z = 2\pi ( 1500 ) \mathrm{Hz}$.

Typical snapshots of the classical field phase distribution in a single temperature quench are shown in Fig.~\ref{fig:ideal-box-protocol}(a),(b) and (c); the three images are taken before (green circle), close to (red circle) and far after (orange circle) the critical point, showing the process of creation and annihilation of vortex-antivortex pairs during the BKT transition. From such distributions, calculated at different times and in many realizations, we extract the correlation function $g^{(1)}$ and the average number of vortices. The results are shown in Fig.~\ref{fig:ideal-box-protocol}(d),(e). The green, red and orange circles in Fig.~\ref{fig:ideal-box-protocol}(e) correspond to the three snapshots of the top row. 

Inspired by \cite{jelic2011quench,Comaron2018}, we investigate whether the system exhibits universal dynamics  in terms of a correlation length $L$. 
To this aim, we first plot the ratio $g^{(1)}/ g^{(1)}_{\mathrm{eq}}$ as a function of $r$  at different times as in the inset of Fig.~\ref{fig:ideal-box-protocol}(d). Following Refs.~\cite{cugliandolo2016,Kulczykowski2017}, we then extract the length $L(t)$ by imposing $(g^{(1)}/g^{(1)}_{\mathrm{eq}})(L(t),t)= 0.2$, and finally we plot  $g^{(1)}/ g^{(1)}_{\mathrm{eq}}$  again, but as a function of the rescaled distance $r/L(t)$. As a result, all curves nicely collapse onto a single one, except at large distances where boundary effects become relevant. This confirms the universal dynamical scaling. We have also checked that the dependence on the value of $g^{(1)}/ g^{(1)}_{\mathrm{eq}}$ chosen for the determination of $L(t)$ is weak and can be neglected. 

\begin{figure*}[t!]
	\centering
	\includegraphics[width=0.488\textwidth]{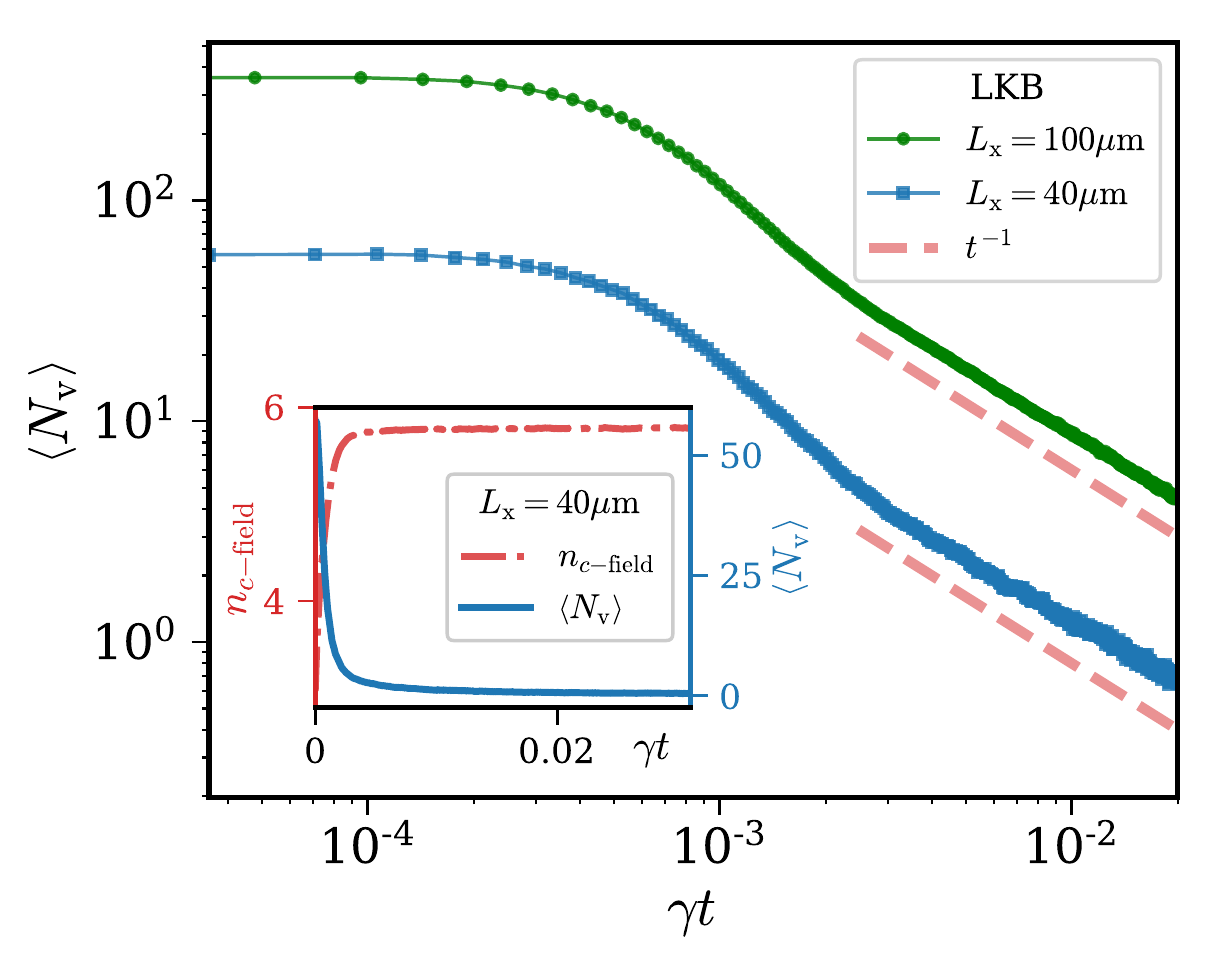}
	\includegraphics[width=0.5\textwidth]{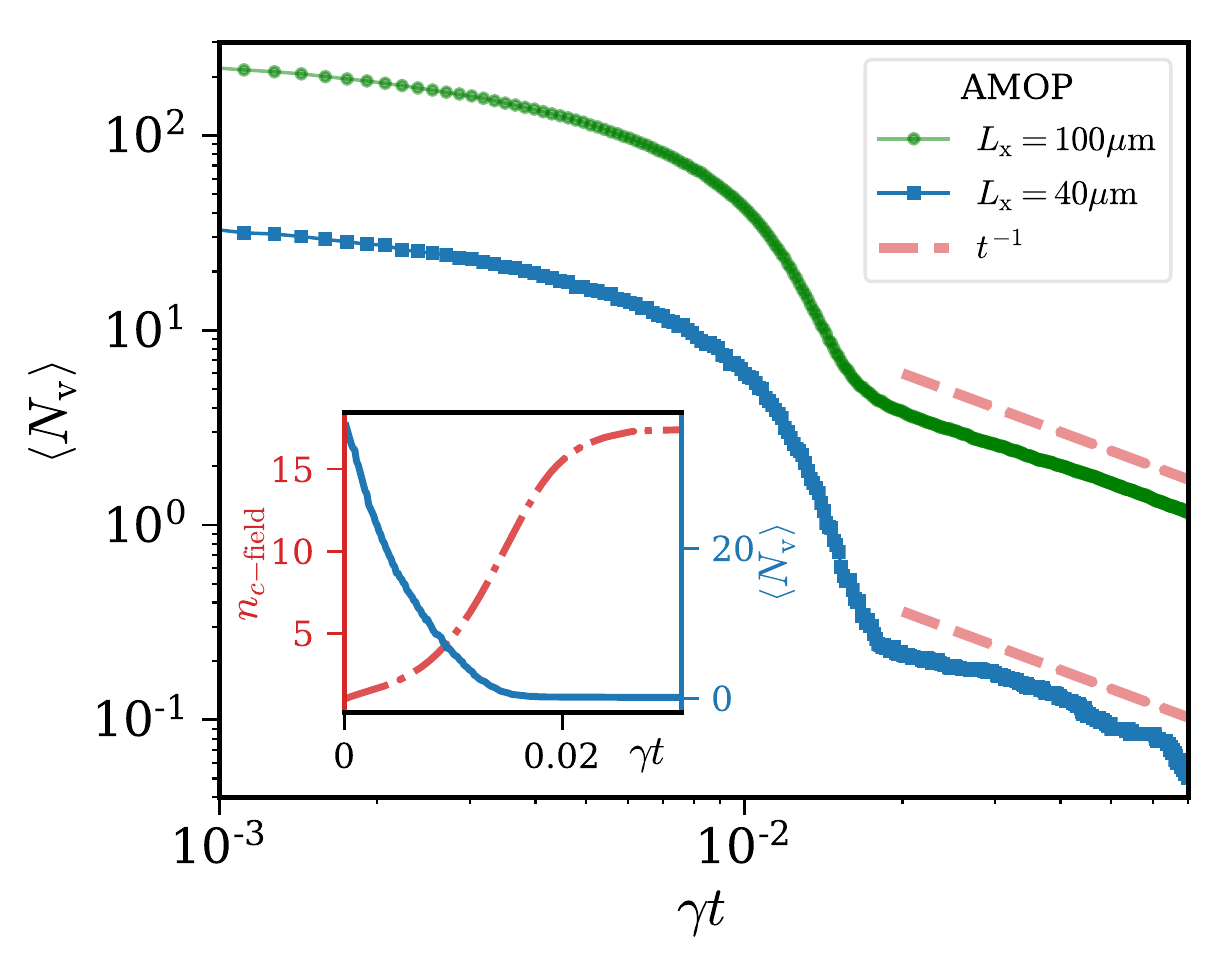}
	\caption{
		Decay of number of vortices following an instantaneous quench of the interaction coupling constant $\tilde{g}$  for two different boxes with $L_x=L_y=40 \mathrm{\mu m}$ (blue {squares and solid line}) and $100  \mathrm{\mu m}$ (green {dots and solid line}), with hard wall boundary conditions, averaged over $\mathcal{N}=400$ stochastic realizations. Results with the parameters of the LKB-Paris and AMOP-Cambridge experimental set-ups are shown on the left and on the right, respectively.  The scaling law $\left< N_\mathrm{v} \right>  \approx t^{-1}$ is shown as a dashed line. Insets: plot in linear scale of the time evolution of average number of vortices $\left< N_\mathrm{v} \right>$ (blue {solid line}) and $c$-field density $n_{\textrm{c-field}}$ (red {dot-dashed line}) in $\mathrm{\mu m^{-2}}$. The key is the same for both insets. In all simulations $\gamma=0.01$.
	}	
	\label{fig:expRel_box}
\end{figure*}

The curve of $L(t)$ is shown in Fig.~\ref{fig:ideal-box-protocol}(e), together with the average number of vortices, as a function of the dimensionless time $\gamma t$.  A key result is that both quantities are found to be related by Eq.~\eqref{eq:relation_L_nv} within the appropriate region where universal scaling is satisfied (grey shaded area in Fig.~\ref{fig:ideal-box-protocol}(e)). Such a window is chosen at late enough times to allow the system to enter the universal regime \cite{note-grey}, but before size effects start to play a role in the growth of $L$ \cite{jelic2011quench,Comaron2018}.
By fitting correlation length and average number of vortices with 
\begin{equation}
		L_{\rm fit} \propto t^{-\beta/2}  \ \ \mathrm{   and   } \ \	\left< N_\mathrm{v} \right>_{\rm fit}  \propto t^{\beta},
\end{equation}
respectively,  we are able to extract the growth exponents for both quantities and we find $\beta \approx -1$.
This is consistent with a value of the dynamical exponent $z= -2/\beta = 2$, as for the case of the planar $XY$ model (with non-conserved order parameter) dynamics, which belongs to the same dynamical universality class.
It is also in agreement with the results obtained within a microcanonical evolution of the system, solved with classical field methods, 
from a many-vortices configuration \cite{Karl2017}. The largest uncertainty in our result for $\beta$ comes from the choice of the parameter $\gamma$ of the SPGPE.  We have performed simulations for different values of $\gamma$ within a decade centred in the value of $\gamma= 0.01$ (which is the range that we think is reasonable for describing our system) and observed $\beta$ to be {compatible with} $z = 2$ within a $10\%$ accuracy. 
The variations of $\gamma$ also reflect on the time values reported in Figs.~\ref{fig:ideal-box-protocol} and~\ref{fig:expRel_box}.
{We have also explicitly verified -- by changing the size and shape of our vortex counting region within the box -- that our numerical extraction of $z=2$ remains unaffected by loss of vortices on the box boundaries.}

In order to make more direct contact with controllable experiments, we also implement a quench in the interaction (i.e., in the coupling constant $\tilde{g}$). Experimentally, varying the value of $\tilde{g}$ is possible by changing the transverse confinement $\omega_z$ (as in the LKB setup), or by tuning the value of the scattering length $a_s$ thanks to the Feshbach resonances \cite{LiChung2013,clark2017collective,HanFu2018}. 
By keeping the control parameters of the reservoir $T_\mathrm{res}$ and $\mu_\mathrm{res}$ fixed, we tune the parameter $\tilde{g}$ such that the system
is suddenly quenched from $\tilde{g}_\mathrm{in}$ to $\tilde{g}_\mathrm{fin}$ with $\tilde{g}_\mathrm{fin} < \tilde{g}_\mathrm{crit} < \tilde{g}_\mathrm{in}$,
with $\tilde{g}_\mathrm{crit}$ defined by Eq.~\eqref{eq:g_crit}.
{Note that fixing $T_\mathrm{res},\mu_\mathrm{res}$ and reducing $\tilde{g}$, causes the critical temperature to increase according to Eq.~\eqref{eq:Tbkt_inf} (distinct to what would happen if one were to instead keep the density constant during the quench). }
The chemical potential $\mu_\mathrm{res} = {16.5  k_\mathrm{B} \mathrm{nK}}$ and the temperature $T_\mathrm{res} = {70 \mathrm{nK}}$ of the reservoir (incoherent region) are chosen in order to have a number of atoms in the box in the order of tens of thousands, consistent with the experimental set-up.
Following Eq.~\eqref{eq:g_crit}, the critical value then reads $\tilde{g}_\mathrm{crit} = 0.18$.

The interaction quench protocol we adopt consists of two stages: first, we prepare our initial {equilibrium, disordered} state $\psi_\mathrm{in} = \psi(\tilde{g}_\mathrm{in},T_\mathrm{res}, \mu_\mathrm{res})$ by evolving Eq.~\eqref{eq:SPGPE} starting from random noise. Then we suddenly quench {to the quasi-ordered state} $\psi_\mathrm{fin} = \psi(\tilde{g}_\mathrm{fin},T_\mathrm{res}, \mu_\mathrm{res})$. 
{
For the initial and final parameter sets ($\tilde{g},T_{\rm res}$ and $\mu_{\rm res}$), we choose values which lie sufficiently above ($\tilde{g}_\mathrm{in}$) and below ($\tilde{g}_\mathrm{fin}$) the critical region, in order to avoid any potential weak dependencies associated with the precise location of the critical region on the implemented energy cutoff.
Specifically, here we use the values 
}
\begin{equation}
\label{eq:range_g_values_gquench}
	\tilde{g}_\mathrm{in} \approx 0.21 \approx 1.19 \tilde{g}_\mathrm{crit} \rightarrow \tilde{g}_\mathrm{fin} \approx 0.12 \approx 0.68 \tilde{g}_\mathrm{crit},
\end{equation}
which, for the parameters of the LKB experiment, would correspond to relaxing the transverse harmonic confinement $\omega_z$ from an initial value $\omega_z~\rightarrow~\omega_z/3$,
where $\omega_z~=~2\pi(7500) \ \mathrm{Hz}$.

Results are shown in Fig.~\ref{fig:ideal-box-protocol}(f),(g). As already done for temperature quenches, we can extract the correlation length $L(t)$ such that the curves of  $g^{(1)}/ g^{(1)}_{\mathrm{eq}}$ collapse onto a universal function of $r/L(t)$. Again we find a time window where the length $L$ and the average number of vortices scale as $t^{-\beta/2}$ and $t^\beta$, respectively, with $\beta \approx -1$ as before.

\subsection{{{Results for experimentally-relevant boxes}}}

In this section we focus on an interaction quench. Contrary to quenching $T$ and $\mu$, the interaction quench is experimentally easier to implement as  it requires only a single parameter to be used to drive the system across the phase transition. Interaction quenches are generally performed in the lab by rapidly changing either the scattering length $a_s$ on times scales~$\approx~100 \ \mathrm{\mu s}$, or the vertical oscillator length $\ell_\perp$ (namely changing the transverse confinement potential  $\omega_z$). 
This protocol is well-known and already adopted for harmonic trap potentials \cite{Fletcher2015} and uniform systems \cite{Ville2018}.

Realistic systems also have finite-size geometries. For this reason in this section we use a box of size $(40 \times 40) \mathrm{\mu m}$, consistent with system sizes of the experiment described in Ref.~\cite{Ville2018}, also displaying results for $(100 \times 100) \mathrm{\mu m}$ to highlight the role of finite-size effects in the smaller systems.
Given the difficulties in measuring $g^{(1)}$ experimentally, we choose to focus the following analysis on the evolution of vortex number, which is also better tractable in numerics, particularly for small system sizes.

First, we repeat the same sudden interaction quench simulation discussed in Fig.~\ref{fig:ideal-box-protocol} (bottom), but for a smaller system where a more realistic hard bound potential box is implemented.
As in the ideal case, we tune the reservoir temperature to $T_\mathrm{res} = {70 \mathrm{nK}}$, and use a modest experimentally-accessible change in transverse confinement, while keeping fixed the value of $\mu_\mathrm{res} = {16.5  k_\mathrm{B} \mathrm{nK}}$.
Thus, the range of the interaction values is as in Eq.~\eqref{eq:range_g_values_gquench}.
Once the system density has reached saturation, we note that the coherent atom fraction reads $n_{\textrm{c-field}}/n \approx 0.81$, with $N \approx 1.2 \times 10^4$.
The numerical cutoff along the whole evolution is fixed by final values of temperature $T_\mathrm{{res}}$, and chemical potential $\mu_\mathrm{res}$.

In addition to the set of parameters of the LKB experiments, we also consider here a rather distinct set of parameters inspired by the AMOP experimental set-up \cite{Campbell2010,Fletcher2015} at Cambridge. 
In this experiment, the choice of $^{39}$K bosons additionally facilitates -- at least in principle -- the control of the interaction strength through a Feshbach resonance. Similarly to the previous case, the temperature and chemical potential of the reservoir set the critical value of the interaction strength $g_\mathrm{{crit}}$. We choose $T = 50 \, \mathrm{nK} $ and {$ \mu = 1.9 \, k_\mathrm{B} \mathrm{nK}$}, corresponding to values of temperature and chemical potential at the end of the evaporation stage of an AMOP experiment  \cite{Hadzibabic}.
Correspondingly, the critical interaction strength becomes $\tilde{g}_\mathrm{crit} \left(\mu,T\right) = 1.83 \times 10^{-2}$, 
so that the quench protocol reads
\begin{equation}
	\tilde{g}_\mathrm{in} = 0.1 \approx 5.46 \, \tilde{g}_\mathrm{crit} \to  \tilde{g}_\mathrm{fin} = 1.83 \times 10^{-3} \approx  0.1\, \tilde{g}_\mathrm{crit}.
\end{equation}
At the end of the simulation the total number of atoms reads $N \approx 29 \times 10^4$, where $n_{c\textrm{-field}}/n \approx 0.98$.

In Fig.~\ref{fig:expRel_box} we show the temporal evolution of the average number of vortices obtained with both sets of parameters. The dashed lines correspond to the universal scaling behavior  $\left< N_\mathrm{v} \right> \propto t^{-1}$.  In order to start observing such a scaling law one has to wait at least a time $t \approx 0.002 \gamma^{-1}$  after the sudden quench for the LKB parameters and $t \approx 0.02 \gamma^{-1}$ for the AMOP parameters, corresponding to $\approx  0.2$~s and $\approx 2$~s, respectively, based on the value of $\gamma=0.01$ used in our simulations. This difference in the timescale for phase ordering to set in can be explained in terms of the different growth rate of coherence that we predict in the two systems with such a value of $\gamma$.  In the inset of  Fig.~\ref{fig:expRel_box} we show the time evolution of the classical field density (red line) and the average number of vortices (blue line) in linear scale.  The classical field density is found to grow faster with the LKB parameters; a factor of approximately ten difference in the growth rate is found by fitting the classical field density with an S-shaped growth curve, as done in Ref. \cite{liu2018dynamical}. This explains the temporal shift of the window where universal scaling is observed in our simulations. This regime occurs roughly at the point when the classical field density appears to approach saturation to its final value, at which however the system has not yet reached its full coherence for the corresponding reservoir parameters. A precise determination of this window in the experiments, however, should require an independent experimental estimate of the coherence growth rate. Nevertheless, time scales in between fractions of a second to ten seconds, as found here, are well within the typical observation time in current experiments. Also, these results appear to be rather independent of the system size.

%% Summary and outlook.
\section{{Conclusions}}

We have performed a detailed analysis of the dynamics following instantaneous temperature and interaction quenches from an incoherent thermal state to a superfluid state below the Berezinskii-Kosterlitz-Thouless phase transition in 2D Bose gases of ultracold atoms.
Considering large boxes with periodic boundary conditions we have demonstrated the self-similarity of correlation functions in the phase-ordering regime, and characterized the evolution of the correlation length and vortex number as a function of time. Both were found to be consistent with a dynamical critical exponent $z=2$, as expected for diffusive dynamics of a system in the 2D $XY$ universality class. Using smaller boxes, we have also shown that realistic geometries, experimentally-accessible interactions and small instantaneous quenches across criticality are likely to facilitate the observation of a regime where $\left< N_\mathrm{v} \right> \propto t^{-1}$ (i.e. $z \approx 2$), thus providing direct measurement of the dynamics and critical exponent $z$ defining the system universality class. 
Our approach is also relevant for investigating the 2D/3D crossover in vortex dynamics, related to recent experiments on vortex clustering and turbulence \cite{Gauthier2018,Johnstone2018,Seo2017}.

\section*{Author Contributions}
P.C. and F.L. contributed equally to this work, undertaking all numerical simulations and analysis and producing the first draft, in direct consultation with F.D. and N.P.P, coordinating the research and leading interpretations. All authors contributed to discussions, final data analysis and interpretations, and the final form of the manuscript.

\section*{{Acknowledgements.}}

We would like to thank C.~Barenghi, T.~Billam, T.~Bland, J.~Dalibard, J.~Schmitt, Z.~Hadzibabic and A.~Groszek for fruitful discussions. 
We acknowledge financial support from the Quantera ERA-NET cofund project NAQUAS through the Consiglio Nazionale delle Ricerche (F.D. and F.L.) and the Engineering and Physical Science Research Council, Grant No. EP/R043434/1 (P.C. and N.P.P.). We acknowledge the Engineering and Physical Science Research Council, Grant No. EP/L504828/1, for DTA support. This work is also supported by Provincia Autonoma di Trento.

%\bibliography{biblio} 

\end{document}